\documentclass[twocolumn,showpacs,preprintnumbers,amsmath,amssymb]{revtex4}

\usepackage{graphicx}
\usepackage{dcolumn}
\usepackage{bm}

\begin{document}

\title{Selective excitation of metastable atomic states by femto- and attosecond laser pulses}

\author{A.~D. Kondorskiy$^{1,2}$}
\author{L.~P. Presnyakov$^{1}$\footnote{Deceased.}}
\affiliation{$^1$ P. N. Lebedev Physical Institute, Leninsky pr., 53, Moscow, 119991, Russia \\
$^2$ Moscow Institute of Physics and Technology, Institutsky per., 9, Dolgoprudny, Moscow region, 141700, Russia}

\author{Yu. Ralchenko}
\affiliation{National Institute of Standards and Technology, Gaithersburg MD 20899-8422, USA}

\date{\today}
 
\begin{abstract}

The possibility of achieving highly selective excitation of low metastable states
of  hydrogen and helium atoms by using short laser pulses with reasonable
parameters  is demonstrated theoretically. Interactions of atoms with the laser
field are studied  by solving the close-coupling equations without
discretization. The parameters of  laser pulses are calculated using different
kinds of optimization procedures. For  the excitation durations of hundreds of
femtoseconds direct optimization of the  parameters of one and two laser pulses
with Gaussian envelopes is used to introduce  a number of simple schemes of
selective excitation. To treat the case of shorter  excitation durations, 
optimal control theory is used and the calculated optimal  fields are
approximated by sequences of pulses with reasonable shapes. A new way  to
achieve selective excitation of metastable atomic states by using sequences of 
attosecond pulses is introduced.

\end{abstract}



\maketitle

\section{\label{sec:intro} Introduction}

Nonlinear laser spectroscopy provides new possibilities to create and study
selectively excited states of quantum systems \cite{Letokhov1977,Hansch1977a}.
Development of the two-photon excitation technique \cite{Hansch1977b} made it
possible to obtain small concentrations of hydrogen in the 2s metastable state.
This was sufficient to carry out precise optical measurements of the hyperfine
structure of 2s state \cite{Kolachevsky2004} and other relativistic and
radiative effects \cite{Fischer2004}. Metastable atoms and atomic ions also play
an important role in excitation and charge transfer processes, even in
high-temperature laboratory and astrophysical plasmas \cite{JPS1985}. Small
concentrations of metastable atoms and atomic ions can significantly affect the
radiative spectra of plasmas because of large cross sections of electronic
capture to the excited states of highly-charged ions during collisions with
the metastable atoms \cite{JPS1985}. This conclusion is based on theoretical and
indirect spectroscopic observations. Unfortunately, direct measurements are
still very difficult due to the low efficiency of modern techniques for the
production of beams of metastable atoms \cite{Gilbody2003}. The above
examples demonstrate the importance of the development of effective methods for
selective excitation of metastable atomic states.

Recent progress in laser techniques as well as in theoretical understanding of
multiphoton processes in atoms makes it possible to introduce novel methods for
effective selective excitation of atomic states by using short laser pulses. A
number of approaches to control population transfer between atomic and molecular
states have been proposed so far. They can be roughly divided into two groups
according to the basic principles. The methods from the first group exploit some
known physical mechanisms \cite{Allen1987, Shore1990,Melinger1992,Band1997, Bergman1998,Unanyan2001,Gong2004,Sola1999,Teranishi1997,Nagaya2001,Zou2005}, 
e.g., it is possible to introduce some \textit{reliable} and \textit{solvable} 
model equations to estimate the parameters of the controlling laser field analytically,
while the methods from the second group are based on the idea of direct variation 
of the laser field to maximize the desired output \cite{Rice2000,Shi1988,
Kosloff1989,Ohtsuki1998,ZBR1998}.

The schemes from the first group ultimately rely on the properties of $\pi$-pulses 
or on adiabatic properties. The controlling laser field is found as an accurate or 
approximate solution of the coupled equations within rotating wave approximation. 
These schemes, in turn, can be classified by the number of laser pulses involved.

Population inversion achieved by using a single $\pi$-pulse represents the
simplest case \cite{Allen1987, Shore1990}. The frequency of the laser should be
in resonance with the transition energy and the control is performed by changing
the pulse area. This method can also be used to control multiphoton transitions.
In that case, the frequency of the laser pulse should be a fraction of the
transition energy so that the energy conservation rules are satisfied. The
advantages of this scheme are that it is simple and that the intensity of the
controlling laser pulse is relatively small. The main disadvantages are that the
efficiency of the scheme is sensitive to the pulse area and that the resonance
conditions should be satisfied accurately, especially in the case of multiphoton
excitation.

The schemes based on adiabatic rapid passage (ARP) rely on adiabatic properties
\cite{Allen1987, Melinger1992, Band1997}. In this case a single chirped laser
pulse is used to establish an adiabatic regime so that the complete population
transfer occurs as a result of system evolution along the initially populated
adiabatic state. The population transfer is not sensitive to the pulse area once
an adiabatic regime is established, so the method is quite robust. However, the
pulse area must be much larger than in the case of $\pi$-pulses. This becomes
critical if the competing ionization or dissociation processes are significant.

Stimulated Raman adiabatic passage (STIRAP) emerged as a very efficient method
to provide almost complete population transfer in a three-level system
\cite{Bergman1998}. In its simplest form, the scheme involves a two-photon Raman
process, in which an interaction with a first (pump) pulse links the initial
state with an intermediate state, which in turn interacts via a second (Stokes)
pulse with a final target state. An important advantage is that almost no
population is placed into the intermediate state, and thus the process is
insensitive to any possible decay from that state. In more complicated cases the
method is used to create a maximal coherent superposition of several states
\cite{Unanyan2001} and complete population transfer via several intermediate
states \cite{Gong2004}. The analog of the STIRAP which relies on adiabatic
properties is called chirped adiabatic passage by two-photon absorption (CAPTA)
\cite{Sola1999}.

Recently an effective scheme based on the idea of periodic chirping was
introduced  \cite{Teranishi1997,Nagaya2001}. Within this scheme, a sequence of
chirped laser  pulses or a laser with periodic chirping is used to establish
multiple crossings  between dressed initial and target levels. Effective laser
control is performed  by manipulating the parameters of these crossings and
adiabatic phase differences  between the two crossings directly. In its simplest
formulation the complete population  transfer can be achieved by using a single
quadratically chirped pulse. Because of  interference between the crossings, the
necessary pulse area of each pulse is much  smaller compared to the ARP; however
the method is still quite robust. These advantages  make the method useful in
the field of wave packet control \cite{Zou2005}.  However, the area of the
pulse is still larger than in the case of control by  non-chirped pulses.

The methods from the second group are based on maximization of the
desired output by a step-by-step adjustment of the controlling field. The main
advantage of these methods is that controlling field search is performed
explicitly, so that complicated systems can be controlled. These methods
can be arranged according to the measure of the controlling parameter space.

In the simplest case, the number of controlling laser pulses and their
envelopes are fixed. Each pulse is described by a few parameters: peak
intensity, central frequency, chirp rate, central time and full width at half
maximum (FWHM). Control is performed by direct optimization of these parameters.
By solving the Schr\"{o}dinger equation, the final population of the desired
state can be determined as a function of the parameters of the controlling
pulses. Thus, well established numerical procedures for maximization of this
function of many variables can be used \cite{Press2002}. It is easy to
demonstrate that all the schemes discussed above can be reproduced by direct
optimization as particular cases. The approach can be considered as an
optimization with respect to a finite set of parameters, i.e., in a 
multi-dimensional space. 

To design the controlling field without any restrictions on its shape, the
optimal control theory (OCT) was developed \cite{Rice2000,Shi1988,
Kosloff1989,Ohtsuki1998,ZBR1998}. It is based on the idea that the controlling
laser pulse should maximize a certain functional. The basic variational
procedure leads to a set of equations for the optimal laser field, which include two
Schr\"{o}dinger equations to describe the dynamics starting from the initial and
target state wave functions. The optimal laser field is given by the imaginary
part of the correlation function of these two wave functions. This system of
equations of optimal control must be solved iteratively in general. The reader
can find a comprehensive review of OCT in \cite{Rice2000}. The main
disadvantage of this method is that in many cases the generated optimal field is
hardly possible to realize in an experiment. The approach can be considered as
optimization with respect to a continious set of parameters, \textit{i.e.}, in
the functional space.

In the case of selective excitation of metastable atomic states by laser
pulses,  a multiphoton interaction is the basic mechanism of the process. Such
effects as  direct ionization \cite{Delone1993, DiMauro1995}, resonant
transitions via   intermediate discrete and continuum states \cite{DiMauro1995,
Freeman1987,  Kondorskiy2001, Kondorskiy2002a, Kondorskiy2003}, above threshold
ionization  \cite{DiMauro1995, Dionissopoulou1995a, Mercouris1996,
Dionissopoulou1995b},  and electron rescattering by the atomic core
\cite{Paulus1994, Paulus1998, Corcum1993,  Shafer1993} are essential. An
accurate treatment of all these effects requires the  exact solution of the
quantum equations to describe the dynamics of the system.  Unfortunately, it is
impossible to introduce some \textit{reliable} and  \textit{solvable} model
equations to estimate the parameters of the controlling  laser field
analytically. Thus, the methods of maximization of desired output  should be
used. 

In the present work selective excitations of the metastable 2s state of hydrogen
and the singlet metastable 1s2s $^{1}$S state of helium are studied using both
the direct optimization of laser parameters and the optimal control theory. To
integrate the time-dependent Schr\"{o}dinger equation, the well
established close-coupling approach based on the properties of time-dependent
integral equations \cite{Kondorskiy2001, Kondorskiy2002a, Kondorskiy2003,
Kondorskiy2002b} is used. This approach is the most suitable for this kind of
spectroscopic calculation. As electron-atom collisions rapidly destroy  atomic
metastable states, a search of the optimal laser parameters is performed under
the requirement that not only high final population of desired metastable state
should be achieved, but also ionization probability should be small
\cite{Kolachevsky2004, Fischer2004}. 

The present paper is organized as follows. In the next section the theoretical 
approaches are summarized, and relations between the results of the optimal
control  calculations and the recently known schemes are demonstrated. In Sec.
\ref{sec:femto}  direct optimization of the parameters of one- and two laser
pulses with Gaussian  envelope is used to introduce a number of simple schemes
for selective excitation  of H(2s) and He(1s2s $^{1}$S). In Sec. \ref{sec:atto}
the optimal control theory  is used to look for other possible controlling laser
fields. A new mechanism of  selective excitation by using attosecond pulses is
introduced. Sec. \ref{sec:concl}  contains a summary.

\section{\label{sec:theory} Theory}

\subsection{\label{subsec:cce} Close-coupling approach}

The time-dependent Schr\"{o}dinger equation for an atom in a laser field is written 
as

\begin{equation}
\left[ i \frac{\partial }{\partial t}-\widehat{H}_{0}+\mathbf{F}(t) \mathbf{d}
(\mathbf{r})\right] \left\vert \Psi (t)\right\rangle =0,
\label{eq:schr_eq}
\end{equation}
where $\widehat{H}_{0}$ is the Hamiltonian of the unperturbed atom, 
$\mathbf{d}(\mathbf{r})$ is the dipole moment and $\mathbf{F}(t)$ is a laser field. 
Atomic units are used throughout the paper, unless otherwise noted.

We employ the close-coupling (CC) method on the basis of the orthogonal and 
normalized unperturbed atomic wave functions of the discrete and continuum states 
and expand the total wavefunction as
\begin{eqnarray}
\left\vert \Psi \left( t\right) \right\rangle = \sum\limits_{\nu } a_{\nu}(t) 
\left\vert \psi _{\nu }\right\rangle 
e^{-iE_{\nu}t}\nonumber \\ 
+ \sum\limits_{\mu }\int b_{\mu E}(t)\left\vert \psi _{\mu E}\right\rangle 
e^{-iEt}dE \label{eq:expan}
\end{eqnarray}
where $\nu $ and $\mu $ are the indices that represent the integer quantum numbers 
of discrete and continuum states respectively. The functions $\left\vert \psi _{\nu }
\right\rangle$ and $\left\vert \psi _{\mu E}\right\rangle$ are the corresponding 
stationary state wavefunctions, $E_{\nu }$ are the energies of the discrete states, 
$E$ stands for the energy of the electron continuum and $a_{\nu}( t)$ and $b_{\mu E}(t)$ 
are unknown coefficients. 

The Hermitian system of coupled equations for the coefficients of the discrete and 
continuum states follows from substituting the expansion Eq.~(\ref{eq:expan}) into 
the time-dependent Schr\"{o}dinger equation Eq.~(\ref{eq:schr_eq}) as
\begin{eqnarray}
i\frac{da_{\nu }\left( t\right) }{dt} = \mathbf{F}(t)\sum\limits_{\nu ^{\prime}}
\mathbf{U}_{\nu \nu ^{\prime }}e^{i\left( E_{\nu }-E_{\nu ^{\prime
}}\right) t}a_{\nu ^{\prime }}\left( t\right) \nonumber \\
 + \mathbf{F}(t)\sum_{\mu }\int \mathbf{U}_{\nu \mu }(E)e^{i\left( E_{\nu }-E\right) 
 t}b_{\mu E}(t)dE, \label{eq:cce1}
\end{eqnarray}
\begin{eqnarray}
i\frac{db_{\mu E}\left( t\right) }{dt} = \mathbf{F}(t)\sum\limits_{\nu } 
\mathbf{U}_{\mu \nu }(E)e^{i\left( E-E_{\nu }\right) t}a_{\nu }\left(
t\right) \nonumber \\
+ \mathbf{F}(t)\sum_{\mu ^{\prime }}\int \mathbf{U}_{\mu \mu ^{\prime }}\left( 
E,E^{\prime }\right) e^{i\left( E-E^{\prime }\right)
t}b_{\mu E^{\prime }}(t)dE^{\prime }. \label{eq:cce2}
\end{eqnarray}
The matrix elements $\mathbf{U}_{\nu \nu ^{\prime }}$, $\mathbf{U}_{\nu \mu }(E)$ 
and $\mathbf{U}_{\mu \mu ^{\prime }}(E,E^{\prime })$ are integrals over 
$\mathbf{r}$-space taken with the atomic dipole moment operator.

The first and the second sums in Eq.~(\ref{eq:cce1}) describe the bound-bound
and free-bound transitions, respectively. These transitions result in a
significant redistribution of the population of discrete states. This in turn
strongly affects the ionization process \cite{Kondorskiy2001, Kondorskiy2002a,
Kondorskiy2003, Kondorskiy2002b}. The first sum in Eq.~(\ref{eq:cce2}) describes
ionization from all the discrete states and the integral term (free-free
transitions) describes a multiphoton inverse bremsstrahlung process (within
quantum mechanical considerations) or rescattering processes (within
quasiclassical considerations). These processes play an important role in
formation of the photoelectronic spectra at high energies \cite{Paulus1994,
Paulus1998, Corcum1993, Shafer1993}. However, since the free-free matrix
elements practically do not affect the discrete state amplitudes $a_{\nu
}\left( t\right) $, it is possible to neglect them for the evaluation of $a_{\nu
}\left( t\right) $. The transitions neglected in this approximation are the
third-order (bound-free-free-bound) ones. The role of the free-free transitions
was carefully investigated in Refs. \cite{Dionissopoulou1995a, Mercouris1996,
Dionissopoulou1995b, Kondorskiy2002a, Kondorskiy2003}, and this assumption was
confirmed to work well.

The close-coupling equations Eqs.~(\ref{eq:cce1}-\ref{eq:cce2}) with the
free-free transitions neglected can be solved by discretizing the continuum or
by employing the recently developed approach based on the properties of the
time-dependent integral equations without any discretization of the continuum
\cite{Kondorskiy2001, Kondorskiy2002a, Kondorskiy2003, Kondorskiy2002b}. As the
nonresonant interaction between the lower and highly excited discrete states is
small, it is possible to adjust the number of discrete states involved in
expansion Eq.~(\ref{eq:expan}) so that the results of the calculations do not
change. Although, strictly speaking, this procedure does not prove the
convergence of the close-coupling approach, it is still widely used in
collisional physics \cite{JPS1985, Lebedev1998}.

In the present study we consider excitation and ionization of a single atomic 
electron in the field of linearly polarized laser pulses.

\subsection{\label{subsec:direct-optim} Methods of direct optimization of 
laser parameters}

In the simplest case the controlling field is assumed to be a sequence of 
laser pulses and is written as
\begin{eqnarray}
F(t)&=&\sum_{j=1}^{N}F_{j}\times f\left( \tau _{j},t-t_{j}\right) \nonumber \\
&\times& \sin \left[\omega _{j}\left( t-t_{j}\right) +\alpha _{j}
\left( t-t_{j}\right) ^{2}\right] \label{eq:gauss_pulses}
\end{eqnarray}
where $f\left( \tau ,\Delta t\right) $ is a fixed envelope function with FWHM
equal to $\tau $ and centered to achieve maximum at time 
difference $\Delta t=0$. In the present study we focus on the laser pulses 
Eq.~(\ref{eq:gauss_pulses}) with a Gaussian shape:
\begin{equation}
f\left( \tau ,\Delta t\right) =\exp \left[ -\frac{4\ln 2}{\tau ^{2}}
\Delta t^{2}\right]. \label{eq:gauss_shape}
\end{equation}

The system is controlled by changing peak amplitudes $F_{j}$, frequencies
$\omega _{j}$, chirp rates $\alpha _{j}$, central times $t_{j}$, and FWHM's
$\tau _{j}$ of the component pulses. By solving the Schr\"{o}dinger equation one
can determine the final population of a selected state as a function of these
parameters. To find the maximum of this function, the conjugate gradient search
method \cite{Press2002} is used here. The gradient of the final population of
the target state with respect to the laser parameters is calculated numerically
using the finite-difference approximation. The optimization starts from some
initial guess parameters. 

As different laser pulse parameters have different dimensions, the parameters 
should be put on a common ground by using dimensionless units, or the
optimizations with respect to different types of parameters should be performed
separately. In the first case the convergence of the optimization procedure 
could depend on the dimensionless units used. Indeed, the efficiency of the
selective  excitation by non-chirped pulse strongly depends on whether or not
the resonant conditions  are achieved, so that by using common atomic units one
should find the maximum  of the function that strongly depends on one group of
arguments (frequencies) and only weakly on the other arguments (intensities,
durations etc.). This  example demonstrates that proper dimensionless units
should be used to ensure  reasonable convergence. Moreover, the derived optimal
parameters could  depend on how that units are defined. 

In the present study the optimizations with respect to  different types of
parameters are performed separately so that each step of the  optimization
procedure should contain a set of optimizations with respect to the  parameters
for all pulses that have the same dimension. Generally, some laser  parameters
should be linked with each other to achieve a maximum of the desired  output. If
the chirping rate is zero, the optimal intensity and duration of each  pulse are
related to each other by some pulse area conservation rule. To avoid this 
uncertainty, we assume the durations of the pulses to be equal. After the
optimal  parameters of the controlling sequence of laser pulses are found, we
can adjust  intensities and durations to fit the specifics of the experimental
technique.

The order of optimizations performed at each step of the direct search affects
the efficiency, convergence and resulting controlling scheme. However, since the
laser field, Eq.~(\ref{eq:gauss_pulses}), has only five different types of
parameters, the number of possible nonequivalent orders of optimization is
limited. Table \ref{tab:table1} presents the correspondence between the
controlling schemes generated by the direct optimization method with different
orders of optimizations performed at each step and previously known schemes. 
Since the last schemes have been established without bound-free transitions 
taken into account the continuum states are not included in the 
test calculations of Table \ref{tab:table1}. The continuum 
states are, however, taken into account at all other calculations, reported in the 
present paper.

The durations of all the pulses are assumed to be equal and do not change during 
optimization. Initial parameters used in all the calculations are 1 $GW/cm^{2}$ 
for intensities of all pulses, frequencies are estimated from the data for the 
unperturbed atom, chirping rates are zero for all the pulses, centers of the 
pulses coincide in the cases 1-4 and are equally distant with the $2\cdot$FWHM 
shift in cases 5 and 6. Our calculations show that number of steps required to 
achieve convergence is about one and a half the number of laser pulses used 
in the scheme.

\begin{table}
\caption{\label{tab:table1} Controlling schemes generated by the direct optimization 
method with different orders of optimizations performed at each step.}
\begin{ruledtabular}
\begin{tabular}{cclc}
\begin{tabular}{c} Case \\ No.\end{tabular} & \begin{tabular}{c} Number \\ of pulses \end{tabular} & 
\begin{tabular}{c} Order of \\ optimizations \end{tabular} & \begin{tabular}{c} Known \\ particular cases \end{tabular} 
\\ \hline \hline
1 & 1 & \begin{tabular}{l} 1. Intensities \\ 2. Frequencies \end{tabular} & $\pi$-pulses 
\\ \hline
2 & 1 & \begin{tabular}{l} 1. Intensities \\ 2. Chirping rates \end{tabular} & ARP \\ 
\hline
3 & 2 & \begin{tabular}{l} 1. Central times \\ 2. Intensities \\ 3. Frequencies \end{tabular} & 
\begin{tabular}{c} Direct excitation \\ or STIRAP \end{tabular} \\ 
\hline
4 & 2 & \begin{tabular}{l} 1. Central times \\ 2. Intensities \\ 3. Chirping rates \end{tabular} & 
\begin{tabular}{c} Direct excitation \\ or CAPTA \end{tabular} \\ 
\hline
5 & Several & \begin{tabular}{l} 1. Frequencies \\ 2. Central times \\ 3. Intensities \end{tabular} & 
\begin{tabular}{l} Periodic sweeping \\ of laser intensity \end{tabular}\\ 
\hline
6 & Several & \begin{tabular}{l} 1. Chirping rates \\ 2. Central times \\ 3. Intensities \end{tabular} & 
\begin{tabular}{c} Periodic sweeping \\ of laser frequency \end{tabular}\\ 
\end{tabular}
\end{ruledtabular}
\end{table}

The procedure of direct optimization of laser parameters can also be used to
calculate  intrinsic parameters of the atomic system. For example, by fixing the
intensity of the  laser and calculating the optimal frequency of selective
excitation of different target  states, the dynamic Stark shifts of these states
can be easily estimated.

\subsection{\label{subsec:oct} Methods of the optimal control theory}

The dynamics of the atom in the laser field is described by the time-dependent 
Schr\"{o}dinger equation (\ref{eq:schr_eq}). The \textit{initial} state wave
function  $\left\vert \Phi _{i}(t)\right\rangle $\ is specified at time $t=0$.
The goal of control  is to design such an external field $\mathbf{F}(t)$\ that
the wave packet calculated  with Eq. (\ref{eq:schr_eq}) up to time $t=T$ is
close enough to the desired  \textit{target} state wave packet $\left\vert
\Phi_{t}(t)\right\rangle $. One of the  most natural and flexible approaches to
design such a field is the optimal control  theory \cite{Rice2000}.
It is based on the idea that the controlling  laser pulse should maximize a
certain functional. The procedure leads to a set of  equations for the optimal
laser field, which include two Schr\"{o}dinger equations to  describe the
dynamics starting from the initial and target state wave packets. The  optimal
laser field is given by the imaginary part of the correlation function of  these
two wave packets. This system of equations of optimal control must be solved 
iteratively in general starting from some initial guess field.

A number of algorithms to realize this idea have been developed 
\cite{Rice2000,Shi1988,Kosloff1989,Ohtsuki1998,ZBR1998}. One of the most effective
is the algorithm by Zhu, Botina and Rabits \cite{ZBR1998} (ZBR algorithm), which is 
developed to solve the following system of the optimal control equations:
\begin{equation}
\left[ i\frac{\partial }{\partial t}-\widehat{H}_{0}+\mathbf{F}(t)\mathbf{d}
(\mathbf{r})\right] \left\vert \phi (t)\right\rangle =0, \quad 
\left\vert \phi(0)\right\rangle =\left\vert \Phi _{i}\right\rangle , \label{eq:zbr-initial}
\end{equation}
\begin{equation}
\left[ i\frac{\partial }{\partial t}-\widehat{H}_{0}+\mathbf{F}(t)\mathbf{d}(\mathbf{r})\right] 
\left\vert \chi (t)\right\rangle =0, \quad 
\left\vert \chi(T)\right\rangle =\left\vert \Phi _{t}\right\rangle , \label{eq:zbr-target} 
\end{equation}
\begin{equation}
\mathbf{F}(t)=-\frac{1}{\gamma }\textrm{Im}\left[ \left\langle \phi (t)\right. \left\vert 
\chi (t)\right\rangle \left\langle \chi (t)\right\vert \mathbf{d}(\mathbf{r})\left\vert 
\phi (t)\right\rangle \right]. \label{eq:zbr-field}
\end{equation}
Here $\gamma$ is a positive parameter chosen to weight the significance of the laser 
energy \cite{ZBR1998}. To integrate Eq. (\ref{eq:zbr-initial}) forward in time and Eq. 
(\ref{eq:zbr-target}) backward in time, we employ the close-coupling method discussed 
in subsection \ref{subsec:cce}.

To launch the ZBR algorithm, an initial field should be specified. In molecular
dynamics control, which is the main area of the application for the ZBR
algorithm, a zero initial field is sufficient for most cases. However, since in
the present study  initial and target states are stationary, a zero initial
field cannot be used.  Instead we use a single laser pulse of Gaussian shape as
an initial field and  generate different optimal control fields by changing its
parameters. Our calculations show that three to five iterations of the ZBR
algorithm are sufficient to achieve the convergence.

\section{\label{sec:femto} Selective excitation of the metastable atomic states 
by femtosecond pulses}

In the present section selective excitations of H(2s) and He(1s2s $^{1}$S) by
one  and two femtosecond laser pulses of Gaussian shape are studied. The case of
chirp  pulses \cite{Allen1987, Melinger1992, Band1997} or periodic chirping 
\cite{Teranishi1997,Nagaya2001} is not considered as the pulse area required is
larger  than in the case of non-chirped pulses. This in turn significantly
increases the  ionization. However, in some cases when the ionization process is
suppressed, periodic chirping can be used to improve the
robustness of the scheme. These will be discussed in a future publication.

\subsection{\label{subsec:femto-single} Selective excitation of H(2s) 
and He(1s2s $^{1}$S) by a single laser pulse}

In the simplest case a two-photon excitation of H(2s) and He(1s2s $^{1}$S) can
be achieved  by using a single laser pulse. The process can be affected by
changing only three  parameters of the laser: frequency, duration, and
intensity. To study this case we calculate  the final populations of the target
metastable states and ionization probabilities as  functions of intensity and
duration of the laser pulse for two-photon excitation of hydrogen  (wavelength
is around 240 nm) and helium (wavelength is around 120 nm). The frequency of the 
laser was optimized for each pair of the arguments to compensate level shifts
due to the  dynamic Stark effect and to maximize the output.

For the case of selective excitation of H(2s) the maximal values of the target
state  population achieved are $18~\%$ to $20~\%$. Unfortunately, the
corresponding ionization  probabilities are large, $27~\%$ to $30~\%$. The
maximal differences between the target state population and ionization
probability are found for the following set of parameters:

\begin{subequations}
\label{eq:hyd_single}
\begin{equation}
\text{FWHM [fs]} = 2.25\times 10^{14} \text{\Large{/}} I\text{ [W/cm}^{2}\text{]},
\end{equation}
\begin{eqnarray}
\omega \text{ [eV]} & = & \text{5.1021 eV} \nonumber \\ & + & 2.34 \times 10^{-15} 
\times I\text{ [W/cm}^{2}\text{]}
\end{eqnarray}
\end{subequations}
where $I$ is the pulse intensity. 
The target state population and ionization probability are found to be $8.4~\%$
 and $4.2~\%$, respectively, for this set of FWHM and $\omega $.
Equations~(\ref{eq:hyd_single}) are obtained  by a fit performed for the range
of intensities from 1 TW/cm$^2$ to 20 TW/cm$^2$.

The efficiency of the process is low because of significance of the
one-photon ionization from the target metastable state. As the photon energy is
about 5.10 eV to 5.14 eV (depending on pulse intensity), this process populates
the continuum states with low energies of about 1.7 eV, so that the
corresponding bound-free transition matrix elements are large.

An opposing situation is observed in the case of selective excitation of He(1s2s $^{1}$S). 
The best values of $65~\%$ for the target state population and $26~\%$ for ionization 
probability are found for the following parameters:
\begin{subequations}
\label{eq:he_single}
\begin{equation}
\text{FWHM [fs]}=7.12\times 10^{15} \text{\Large{/}} I\text{ [W/cm}^{2}\text{]},
\end{equation}
\begin{eqnarray}
\omega \text{ [eV]} & = & \text{10.3075 eV} \nonumber \\ & 
+ & 1.54 \times 10^{-16} \times I\text{ [W/cm}^{2}\text{]}.
\end{eqnarray}
\end{subequations}
Eqations~(\ref{eq:he_single}) are obtained by a fit performed for the range of 
intensities from 50 TW/cm$^2$ to 400 TW/cm$^2$.

The photon energy is about two times higher then in the case of hydrogen (10.3
eV to 10.4 eV, depending on pulse intensity). The process is effective.
One-photon ionization from the target metastable state populates the continuum
states with energies of about 6.3 eV and corresponding bound-free transition
matrix elements are small, so that ionization does not undermine the process.

\subsection{\label{subsec:femto-two} Selective excitation of H(2S) by two laser pulses}

\begin{figure}[t]
\centering
\includegraphics[width=0.9\linewidth]{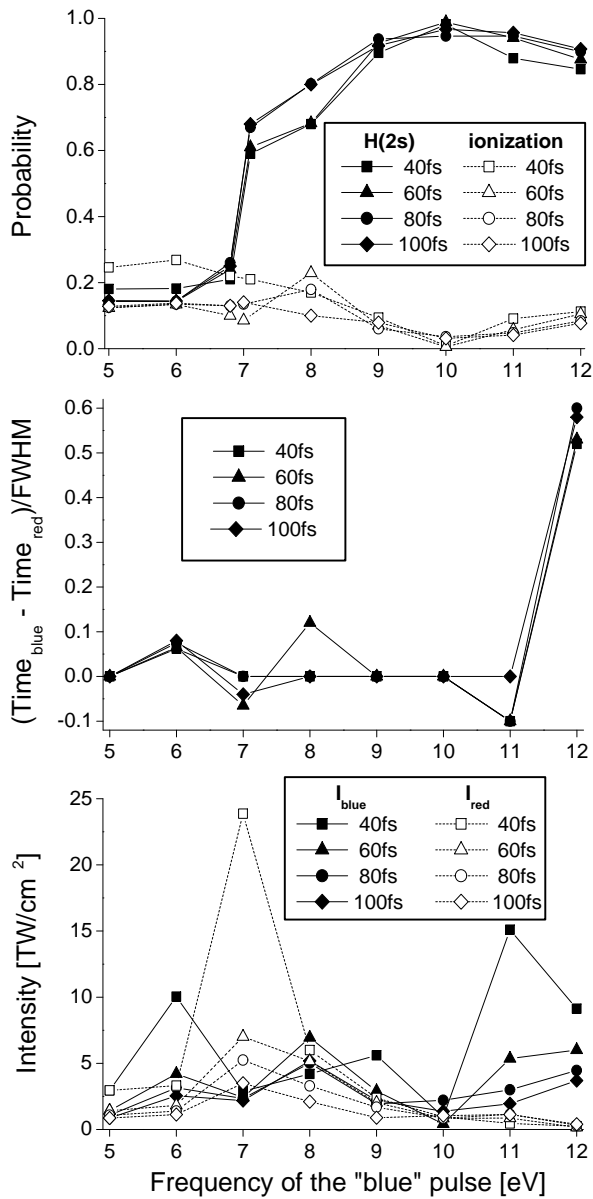}
\caption[]{Selective excitation of H(2S) by two laser pulses: (top) final populations of 
2s metastable state and ionization probability, (middle) time shift between the ``blue'' 
and ``red'' pulses and (bottom) intensities of the pulses given as functions of 
frequency of the ``blue'' pulse.}
\label{fig:1}
\end{figure}

To improve the efficiency of selective excitation of the metastable state of
hydrogen,  we increase the number of controlling parameters by introducing two
controlling laser  pulses with different frequencies instead of a single pulse.
For simplicity, we call  the pulse with lower frequency as ``red'' pulse and the
pulse with higher frequency as  ``blue'' pulse. Frequencies of the blue and
the red pulses should be linked  to fit the energy difference between the
ground and the metastable states. As optimal intensity and duration of each pulse 
are approximately related to each other by a pulse envelope area conservation rule, 
in the present study we fix durations of the pulses to be equal.

Figure \ref{fig:1} presents the final population of the metastable state and
ionization  probability of hydrogen as functions of frequency of the blue
pulse for different  durations of the pulses with the frequency of the red pulse,
intensities and central  times of both pulses adjusted to maximize the
population of the metastable state.

If the frequency of the red pulse becomes lower than the threshold of the
one-photon ionization of the 2s metastable state, the efficiency of the
selective excitation grows dramatically. This is caused by an abrupt decrease of
ionization rate, which makes it possible to apply more intense laser pulses and
achieve higher selective excitation without increasing the ionization.
Unfortunately, for hydrogen the one-photon ionization threshold for the 2s state
is equal to one third of the energy difference between the 1s and 2s states, and
the simple case, when the blue pulse is produced from the single red
pulse by the second harmonic generation, is still not good enough. Acceptable
results, however, can be obtained for the frequencies of the blue pulse
slightly higher and frequencies of the red pulse slightly lower than the
threshold values. An interesting observation comes from the analysis of the time
shift between the two optimized controlling pulses. If the frequencies of the pulses
are far from being in resonance with some bound state,  their centers coincide.
However, if the frequency of the blue pulse is close to the frequency of 3p
$\rightarrow$ 1s transition, the centers of the blue and the red pulses
separate. In this case the direct two-photon excitation process transforms into
STIRAP \cite{Bergman1998} (Table \ref{tab:table1}, case 3).

Consider three examples in more detail:

\textbf{\textit{Example 1.}} \textit{Final populations: 2s - 59 \%, continuum -
21 \%.  Parameters of the pulses. Freqencies: blue - 7.12786 eV, red -
3.13629 eV;  Intensities: blue - $5.8 \times 10^{12}$ $W/cm^{2}$, red -
$1.2 \times 10^{13}$ $W/cm^{2}$;  FWHMs are 40 fs for both pulses; Pulse centers
coincide.} The frequency of the  red pulse  is below but close to the
threshold of the one-photon ionization of the H(2s) state. Although the 
frequencies are adjusted to avoid resonances with highly excited discrete
states, 7p and  8p states are  populated at the level of 9 \% and 3 \%
respectively. 

\textbf{\textit{Example 2.}} \textit{Final populations: 2s - 92 \%, continuum -
7 \%.  Parameters of the pulses. Freqencies: blue - 9.09943 eV, red -
1.09933 eV;  Intensities: blue - $4.8 \times 10^{12}$ $W/cm^{2}$, red -
$1.6 \times 10^{12}$  $W/cm^{2}$; FWHMs: blue - 40 fs, red - 100 fs;
Pulse centers coincide.}  In this case durations of the pulses are different and
efficiency of the process is  less sensitive to the shifts between the pulse
centers.

\textbf{\textit{Example 3.}} \textit{Final populations: 2s - 91 \%, continuum -
8 \%.  Parameters of the pulses. Freqencies: blue - 12.0878 eV, red -
1.88656 eV;  Intensities: blue - $3.7 \times 10^{12}$ $W/cm^{2}$, red -
$4.0 \times 10^{11}$  $W/cm^{2}$; FWHMs are 100 fs for both pulses; blue
pulse is delayed for 58 fs.}  The frequencies of the pulses are close to be in
resonance with 3p state. In this case  the red pulse comes before the
blue pulse. This represents the STIRAP process  \cite{Bergman1998}.

\section{\label{sec:atto} Selective excitation of the metastable atomic states 
by attosecond pulses}

\begin{figure}[t]
\centering
\includegraphics[width=0.9\linewidth]{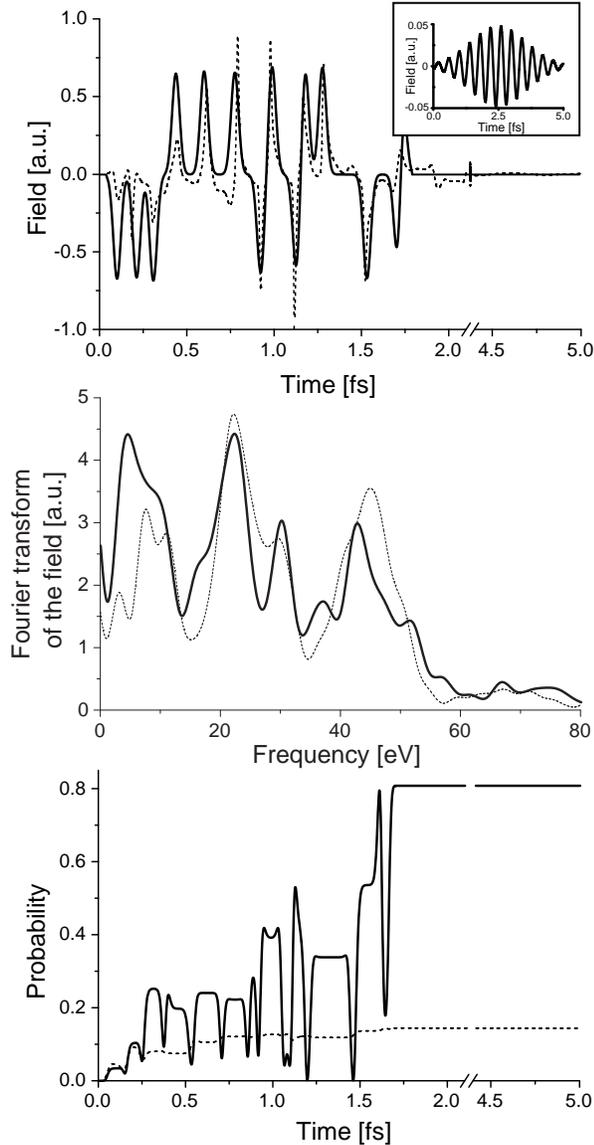}
\caption[]{The optimal control field generated for selective excitation of 
the 1s2s $^{1}$S state of helium (top), its Fourier transform (middle) and 
time variation of the metastable state population and ionization probability 
of the controlled atom (bottom). Top and middle: dashed line is the OCT result, 
solid line is the fitted sequence of attosecond pulses. Insert: initial guess 
field. Bottom: solid line is the population of the 1s2s $^{1}$S state, dashed 
line is the ionization probability.}
\label{fig:2}
\end{figure}

\begin{figure}[t]
\centering
\includegraphics[width=0.9\linewidth]{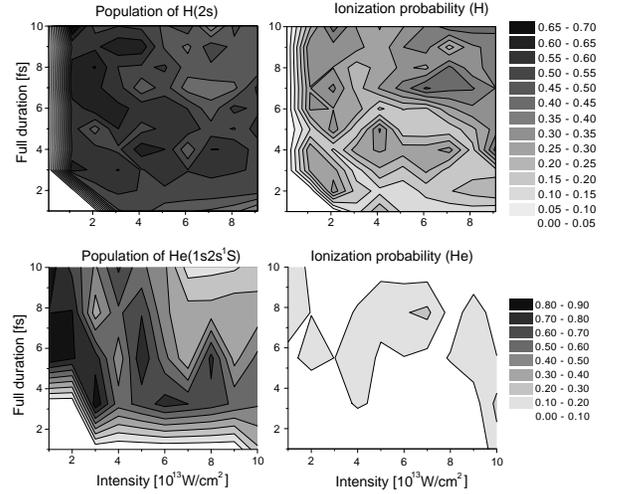}
\caption[]{Final populations of the metastable states and ionization probabilities 
obtained with the optimal field generated from a single Gaussian laser pulse given 
as a functions of the parameters of initial guess pulse. Top line: Results obtained 
for selective excitation of H(2s) with the frequency of initial guess pulse is 10.2 
eV. Bottom line: Results obtained for selective excitation of H(1s2s $^{1}$S) with 
the frequency of initial guess pulse is 10.3 eV.}
\label{fig:3}
\end{figure}

In the previous sections the controlling laser field was assumed to be composed 
of a few femtosecond laser pulses with Gaussian envelopes. To study other 
possibilities to achieve selective excitation we use the optimal control theory, 
as described in Sec. \ref{subsec:oct}. As the controlling schemes introduced in 
the previous section for long durations can hardly be improved, we focus on the 
case of short excitation times.

A typical example is shown in Fig. \ref{fig:2}. It presents the optimal control 
field generated by the ZBR algorithm for selective excitation of 1s2s $^{1}$S
state  of helium (top) and the time variation of population of the target
metastable  state and ionization probability during the interaction (bottom).
The calculations  are performed starting from the Gaussian pulse of intensity 80
TW/cm$^{2}$,  frequency 10.3 eV and FWHM 2.5 fs (full duration is 5 fs), which is
shown in the  insert of Fig. \ref{fig:2}. The weight parameter is taken to be
$\gamma = 0.1$.  The final population of the target state is $77~\%$ with
ionization probability $7~\%$. The convergence is achieved in four iterations.

During optimization the laser field completely transforms into a sequence of 
attosecond pulses. The main perturbations finish after 2 fs, so the duration of 
active control is also changed. Contrary to the case of controlling femtosecond 
pulses, the state populations now change abruptly. Physically, the sequence of 
attosecond laser pulses affects the atomic electrons as a series of pushes. Each 
push displaces the electrons for a small distance so they can not move far from 
the atomic core, and this suppresses ionization.

To simplify the optimal control field obtained with the ZBR algorithm we approximate 
it by a sequence of Gaussians:
\begin{equation}
F(t)=\sum_{j=1}^{N}\frac{A_{j}}{\sqrt{\beta \pi }}\exp \left[ -\beta \left(t-t_{j}\right) 
\right]. \label{eq:atto_approx}
\end{equation}
If the duration of each pulse is short enough, the momentum supplied to the
electrons  by each attosecond pulse is given by the pulse area $A_{j}$ and does
not depend on  $\beta$. To calculate the parameters ${ A_{j} }$ and ${ t_{j} }$
we roughly approximate  the controlling field by sequence (\ref{eq:atto_approx})
and use the method described  in Sec. \ref{subsec:direct-optim} to optimize ${
A_{j} }$ and ${ t_{j} }$. The resulting  sequence of attosecond pulses is presented
in Fig. \ref{fig:2} (top) as a bold line. The  calculation is performed with the width
parameter $\beta = 0.04$ (FWHM = 20 as).  The final population of the target
state and the ionization probability for that field  are $81~\%$ and $14~\%$,
respectively.

Figure \ref{fig:3} presents final populations of the metastable states and ionization 
probabilities obtained with the optimal fields generated from femtosecond pulses of 
different intensities and durations. Each point of the plot represents an
effective controlling sequence of attosecond pulses. One can see that the 
results depend on the parameters of the initial guess pulse in a very complicated way. 
The parameters for some effective controlling sequences of attosecond pulses 
Eq. (\ref{eq:atto_approx}) calculated by optimal fitting of the parameters 
${ A_{j} }$ and ${ t_{j} }$ are presented in the Table \ref{tab:table2}.

Recently, q number of schemes to generate trains of attosecond laser pulses 
have been proposed and realized experimentally (see review
\cite{Agostini2004}).  The basic idea is the production of a comb of equidistant
frequencies in the  spectral domain with controlled relative phases. As a result
trains of  x-ray laser pulses of attosecond duration are obtained. The shapes
and carrier  frequency can be modified by eliminating certain harmonics. Also,
waveforms  containing optical bursts approaching one cycle have been
successfully  synthesized by superposing five phase-controlled sidebands
\cite{Sokolov2001}.  Trains of non-modulated attosecond pulses could be also
generated by laser-plasma interaction in the relativistic regime (see
\cite{Pirozhkov2006} and references therein.)

To generate sequence of pulses close to ones shown in Fig. 2, the spectral
components  with the frequencies $7.33$ eV, $22.0 \approx 3 \times 7.3$ eV,
$29.3\approx 4 \times  7.3$ eV and $44.0 \approx 6 \times 7.3$ eV should be
syncronized. Four main harmonics taken from equidistant spectral modes should
be superposed. Thus, for principal frequency  of 7.33 eV one should take 1st, 3rd, 4th
and 6th harmonics; for principal frequency  of 3.66 eV 2nd, 6th, 8th and 12th
harmonics, respectively, and so on.

\begin{table*}
\caption{\label{tab:table2} Parameters of some effective controlling sequences 
of attosecond pulses Eq. (\ref{eq:atto_approx}) calculated by optimal fitting of 
the optimal control fields. Column 1: target atomic state. Colunms 2-4: parameters 
of the initial guess femtosecond pulse. Columns 5-6: final target state populations 
and ionization probabilities. Colunms 7-8: Parameters of the sequencies of the 
attosecond pulses.}
\begin{ruledtabular}
\begin{tabular}{cccccccc}

\begin{tabular}{c} Target \\ state \end{tabular} & \begin{tabular}{c} $\omega_{ini}$ \\ $[eV]$ \end{tabular} & \begin{tabular}{c} $\text{FWHM}_{ini}$ \\ $[fs]$ \end{tabular} & \begin{tabular}{c} $I_{ini}$ \\ $[\text{W/cm}^{2}]$ \end{tabular} & $P_{target}$ & $P_{ion}$ & \begin{tabular}{c} $t_{j}$ \\ $[fs]$ \end{tabular} & \begin{tabular}{c} $A_{j}$ \\ $[a.u.]$ \end{tabular} \\ \hline \hline

H(2s) & 10.2 & 5 & $1.0 \times 10^{13}$ & 0.67 & 0.31 &
\begin{tabular}{l} 0.05, 0.32, 0.76, 1.33, 1.51, \\ 1.85, 2.02, 2.13, 2.36, 2.41, \\ 2.73 \end{tabular} & 
\begin{tabular}{l} -0.50, -0.44, -0.40, -0.93, -0.44,  \\ -0.84, -0.83, -0.93, -0.67, 0.46, \\ 0.95 \end{tabular} \\ \hline

H(2s) & 10.2 & 3.5 & $1.5 \times 10^{13}$ & 0.65 & 0.29 &
\begin{tabular}{l} 0.05, 0.69, 1.15, 1.90, 2.27, \\ 2.78, 3.03, 3.23, 3.45, 3.50 \end{tabular} & 
\begin{tabular}{l} 0.29, 0.47, 0.30, -0.33, -0.31, \\ -0.44, 0.31, -0.46, 0.61, 0.57 \end{tabular} \\ \hline

H(2s) & 10.2 & 4.0 & $2.3 \times 10^{13}$ & 0.63 & 0.29 &
\begin{tabular}{l} 0.05, 0.40, 0.45, 0.77, 0.94, \\ 1.19, 1.34, 1.56, 1.86, 1.93, \\ 2.08, 2.30, 2.45 \end{tabular} & 
\begin{tabular}{l} 0.44, 0.41, 0.4, 0.57, 0.41, \\ 0.44, 0.33, 0.26, 0.61, 0.58, \\ 0.49, -0.34, -0.81 \end{tabular} \\ \hline

H(2s) & 10.2 & 2.0 & $7.5 \times 10^{13}$ & 0.62 & 0.25 &
\begin{tabular}{l} 0.05, 0.35, 0.68, 1.03, 1.22, \\ 1.30, 1.33, 2.0, 2.45  \end{tabular} & 
\begin{tabular}{l} -0.09, -0.48, -0.34, -0.29, -0.7, \\ -0.23, -0.31, -0.71, -1.43  \end{tabular} \\ \hline

H(2s) & 5.1 & 5.0 & $1.0 \times 10^{12}$ & 0.70 & 0.20 &
\begin{tabular}{l} 0.05, 0.25, 0.80, 1.17, 1.50, \\ 1.55, 1.64, 2.14, 2.25, 2.32, \\ 2.70, 2.76 \end{tabular} & 
\begin{tabular}{l} 0.24, -0.43, -0.29, -0.43, -0.44, \\ -0.44, -0.29, -0.52, -0.57, -0.59, \\ 0.47, 0.43 \end{tabular} \\ \hline

H(2s) & 5.1 & 3.5 & $1.0 \times 10^{12}$ & 0.67 & 0.21 &
\begin{tabular}{l} 0.05, 0.28, 0.55, 0.71, 1.31, \\ 0.96, 1.09 \end{tabular} & 
\begin{tabular}{l} -0.2, 0.13, 0.37, -1.4, -0.57, \\ 0.39, 0.61 \end{tabular} \\ \hline

H(2s) & 5.1 & 2.5 & $1.0 \times 10^{12}$ & 0.65 & 0.25 &
\begin{tabular}{l} 0.05, 0.46, 0.87, 1.14, 1.22, \\ 1.42, 1.58, 1.66 \end{tabular} & 
\begin{tabular}{l} 0.40, 0.33, 0.19, 0.57, 0.51, \\ 0.52, -0.21, -0.81 \end{tabular} \\ \hline

He(1s2s $^{1}$S) & 10.3 & 5.0 & $8.0 \times 10^{13}$ & 0.81 & 0.14 &
\begin{tabular}{l} 0.05, 0.16, 0.25, 0.38, 0.54, \\ 0.71, 0.86, 0.92, 1.06, 1.11, \\ 1.21, 1.45, 1.63, 1.65 \end{tabular} & 
\begin{tabular}{l} -0.71, -0.71, -0.73, 0.70, 0.71, \\ 0.71, -0.69, 0.74, -0.69, 0.73, \\ 0.73, -0.71, -0.77, 0.66 \end{tabular} \\ \hline

He(1s2s $^{1}$S) & 10.3 & 5.0 & $5.0 \times 10^{13}$ & 0.75 & 0.10 &
\begin{tabular}{l} 0.05, 0.11, 0.23, 0.57, 0.73, \\ 0.89, 0.95, 1.07, 1.16, 1.27, \\ 1.32, 1.44, 1.48, 1.64, 1.67 \end{tabular} & 
\begin{tabular}{l} -0.70, -0.66, -0.61, 0.51, -0.54, \\ -0.6, -0.69, 0.86, -0.71, 0.65, \\ -0.76, 0.66, -0.73, -0.71, 0.69 \end{tabular} \\ \hline

He(1s2s $^{1}$S) & 10.3 & 1.0 & $2.0 \times 10^{14}$ & 0.78 & 0.06 &
\begin{tabular}{l} 0.05, 0.17, 0.24, 0.36, 0.45, \\ 0.51, 0.63, 0.73, 0.76 \end{tabular} & 
\begin{tabular}{l} -0.56, -0.53, -0.89, -0.59, -0.94, \\ 0.81, -0.66, 0.70, -0.79 \end{tabular} \\ \hline

He(1s2s $^{1}$S) & 10.3 & 6 & $1.0 \times 10^{13}$ & 0.79 & 0.10 &
\begin{tabular}{l} 0.05, 0.25, 0.33, 0.41, 0.57, \\ 0.72, 0.78, 0.90, 0.98, 1.05, \\ 1.15 \end{tabular} & 
\begin{tabular}{l} 1.0, 0.88, 0.43, 0.36, 0.51, \\ 0.66, 0.71, 0.74, 0.80, -0.78, \\ 0.90 \end{tabular} \\ \hline

\end{tabular}
\end{ruledtabular}
\end{table*}

\section{\label{sec:concl} Conclusions}

In the present study the controlling field search is performed using a well
established close-coupling approach for solving the time-dependent
Schr\"{o}dinger equation, so that all essential multiphoton effects are treated
accurately. It is demonstrated that optimization procedures provide an effective
technique to design laser fields for selective excitation of metastable atomic
states. Not only do they reproduce the recently developed schemes of laser control
as particular cases but also they introduce new ones. 

An efficient selective excitation of H(2s) and He(1s2s $^{1}$S) can be achieved 
by using one and two femtosecond laser pulses. Frequencies of the pulses should 
be adjusted properly to suppress single-photon ionization from the target
metastable  state. Optimal populations of the atomic states and the
corresponding parameters  of the laser pulses are calculated as functions of
preferable frequencies and durations of the pulses. These make it possible to
choose the parameters of the  controlling laser pulses that fit the capabilities
of the present experimental techniques.

A new way to achieve selective excitation of metastable atomic states by using 
sequences of attosecond pulses is introduced. This is important because of recent 
progress in production of attosecond pulses. While in the present study the 
parameters of the pulses are calculated using the optimal control theory, the 
direct optimization of high harmonics can be used to control the process in the future.

\begin{acknowledgments}

We wish to thank Prof. I. I. Sobel'man, Prof. N. B. Delone, Prof. A. N.
Grum-Grzhimailo,  Prof. H. Nakamura, Dr. N. N. Kolachevsky and Dr. A. A. Narits
for useful discussions. This work is supported in part by the Russian Foundation 
for Basic Research (project 06-02-17089) (A. K.) and the Office of Fusion
Energy Sciences of the U.S. Department of Energy (Yu. R.).

\end{acknowledgments}


\begin{thebibliography}{0}

\bibitem{Letokhov1977} V. S. Letokhov and V. P. Chebotaev, {\it Nonlinear laser spectroscopy} (Springer-Verlag, Berlin-New, York 1977).

\bibitem{Hansch1977a} T. W. H\"{a}nsch, {\it Nonlinear Spectroscopy}, edited by N. Bloembergen (Amsterdam, New-Holland 1977).

\bibitem{Hansch1977b} T. W. H\"{a}nsch, {\it Laser Spectroscopy III, Springer Series in Optical Sciences 7} edited by J. L. Hall and S. L. Carlsten, (Springer, Berlin-New York 1977).

\bibitem{Kolachevsky2004} N. Kolachevsky, M.Fischer, S. G. Karshenboim and T. W. H\"{a}nsch, Phys. Rev. Lett. {\bf 92}, 033003, (2004).

\bibitem{Fischer2004} M. Fischer {\it et al}, Phys. Rev. Lett. {\bf 92}, 230802, (2004).

\bibitem{JPS1985} R. K. Janev, L. P. Presnyakov and V. P. Shevelko, {\it Physics of Highly Charged Ions} (Springer Verlag, Berlin 1985).

\bibitem{Gilbody2003} H. B. Gilbody, {\it private communication} (2003).

\bibitem{Allen1987} L. Allen and J. H. Eberly, {\it Optical Response and Two-Level Atoms} (Dover, New York, 1987).

\bibitem{Shore1990} B. W. Shore, {\it The Theory of Coherent Atomic Excitation} (Wiley, New York 1990).

\bibitem{Melinger1992} J. S. Melinger, S. R. Gandhi, A. Hariharan, J. X. Tull and W. S. Warren, Phys. Rev. Lett. {\bf 68}, 2000 (1992).

\bibitem{Band1997} Y. B. Band and P. S. Julienne, J. Chem. Phys. {\bf 97}, 9107 (1997).

\bibitem{Bergman1998} K. Bergman, H. Theuer and B. W. Shore, Rev. Mod. Phys. {\bf 70}, 1003 (1998).

\bibitem{Unanyan2001} R. G. Unanyan, B. W. Shore and K. Bergmann, Phys. Rev. A {\bf 63}, 043401 (2001).

\bibitem{Gong2004} J. Gong and S. A. Rice, Phys. Rev. A {\bf 69}, 063410 (2004).

\bibitem{Sola1999} I. R. Sol\'a, V. S. Malinovsky, B. Y. Chang, J. Santamaria and K. Bergmann, Phys. Rev. A. {\bf 59}, 4494 (1999).

\bibitem{Teranishi1997} Y. Teranishi and H. Nakamura, Phys. Rev. Lett. {\bf 81}, 2032 (1998).

\bibitem{Nagaya2001} K. Nagaya Y. Teranishi and H. Nakamura, {\it Advances in Multiphoton Progress and Spectroscopy}, edited by R. J. Gordon and Y. Fujimura (World Scientific, Singapore, 2001), Vol. 14.

\bibitem{Zou2005} S. Zou, A. Kondorskiy, G. Mil'nikov and H. Nakamura, J. Chem. Phys. {\bf 122}, 084112 (2005).

\bibitem{Rice2000} S. A. Rice and M. Zhao, {\it Optical Control of Molecular Dynamics} (John-Willey \& Sons, 2000).

\bibitem{Shi1988} S. Shi, A. Woody and H. Rabitz, J. Chem. Phys. {\bf 88}, 6870 (1988).

\bibitem{Kosloff1989} R. Kosloff, S. Rice, P. Gaspard, S. Tersigni and D. Tannor, Chem. Phys. {\bf 139}, 201 (1989).

\bibitem{Ohtsuki1998} Y. Ohtsuki, H. Kono and Y. Fujimura, J. Chem. Phys. {\bf 109}, 9318 (1998).

\bibitem{ZBR1998} W. Zhu, J. Botina and H. Rabitz, J. Chem. Phys. {\bf 108}, 1953 (1998).

\bibitem{Press2002} W. H. Press, S. A. Teukolsky, W. T. Vetterling and B. P. Flannery, {\it Numerical Recepies in Fortran}; {\it Numerical Recepies in C}; {\it  Numerical Recepies in C++} (Cambridge University Press, 2001-2002).

\bibitem{Delone1993} N. B. Delone, {\it Basis of Interaction of Laser Radiation with Matter} (Gif-Sur-Yvette Cedex-France, Editiones Frontieres 1993).

\bibitem{DiMauro1995} L. F. DiMauro and P. Agostini, Adv. At. Mol. Opt. Phys. {\bf 35}, 79 (1995).

\bibitem{Freeman1987} R.R. Freeman {\it et al}, Phys. Rev. Lett., {\bf 59}, 1092 (1987).

\bibitem{Kondorskiy2001} A. D. Kondorskiy and L. P. Presnyakov, J. Phys. B: At. Mol. Opt. Phys. {\bf 34}, L663 (2001).

\bibitem{Kondorskiy2002a} A. D. Kondorskiy and L. P. Presnyakov, Laser Phys. {\bf 12}, 449 (2002).

\bibitem{Kondorskiy2003} A. D. Kondorskiy and L. P. Presnyakov, Proc. SPIE {\bf 5228}, 394 (2003).

\bibitem{Kondorskiy2002b} A. Kondorskiy, H. Nakamura, Phys. Rev. A {\bf 66}, 053412 (2002).

\bibitem{Dionissopoulou1995a} S. Dionissopoulou, Th. Mercouris, A. Lyras, Y. Komninos and C. A. Nicolaides, Phys. Rev. A {\bf 51}, 3104 (1995).

\bibitem{Mercouris1996} T. Mercouris {\it et al}, J. Phys. B: At. Mol. Opt. Phys. {\bf 29}, L13 (1996).

\bibitem{Dionissopoulou1995b} S. Dionissopoulou, Th. Mercouris, A. Lyras and C. A. Nicolaides, Phys. Rev. A {\bf 55}, 4397 (1997).

\bibitem{Paulus1994} G. G. Paulus {\it et al}, J. Phys. B: At. Mol. Opt. Phys. {\bf 27}, L703, (1994).

\bibitem{Paulus1998} G. G. Paulus {\it et al}, Phys Rev. Lett. {\bf 80}, 484 (1998).

\bibitem{Corcum1993} P. Corkum, Phys. Rev. Lett., {\bf 71}, 1994 (1993).

\bibitem{Shafer1993} K. J. Schafer, Baorui Yang, L. F. DiMauro and K. C. Kulander, Phys. Rev. Lett. {\bf 70}, 1599 (1993).

\bibitem{Lebedev1998} V. S. Lebedev and I. L. Beigman, {\it Physics of Highly Excited Atoms and Ions}, (Springer, Berlin-Heidelberg, 1998).

\bibitem{Agostini2004} P. Agostini and L. F. DiMauro, Rep. Prog. Phys., {\bf 67}, 813 (2004).

\bibitem{Sokolov2001} A. V. Sokolov {\it et al}, Phys. Rev. A, {\bf 63}, 051801 (2001); Phys. Rev. Lett, {\bf 87}, 033402 (2001);

\bibitem{Pirozhkov2006} A. S. Pirozhkov {\it et al}, Physics of Plasmas, {\bf 13}, 013107 (2006).






\end{thebibliography}
\end{document}